\documentclass{emulateapj}
\usepackage{color}
\slugcomment{Activity published in \href{http://astroedu.iau.org/activities/street-lights-standard-candles/}{astroEDU, 1535 (2016)}}
\shorttitle{Street lights as standard candles}
\shortauthors{M. P\"ossel}

\begin{document}

\begin{abstract}
Astronomers measure cosmic distances to objects beyond our own galaxy using standard candles: objects of known intrinsic brightness, whose apparent brightnesses in the sky are then taken as an indication of their distances from the observer. In this activity, we use street lights and a digital camera to explore the method of standard candles as well as some of its limitations and possible sources of error.\end{abstract}

\title{Street lights as standard candles: A student activity for understanding
astronomical distance measurements}
\author{Markus P\"ossel}
\affil{Haus der Astronomie and Max Planck Institute for Astronomy\\
K\"onigstuhl 17, 69117 Heidelberg}
\email{E-mail: poessel@hda-hd.de}
\maketitle

\section{Astronomical distance measurements}
The astronomer's typical role is that of a distant observer, trying to reconstruct an object's physical properties from afar. The key to inferring such properties is knowledge of the object's distance. After all, bright objects can look dim when seen from afar, and large objects can look small. Only when an object's distance to us is known can we relate observational data to physical properties: The object's angular diameter to its physical size, or the amount of light we receive from an object to its intrinsic brightness.

Unfortunately, estimating cosmic distances is a difficult task. As no single method can cover the whole gamut of astronomical distances, astronomers have constructed a {\em cosmic distance ladder} that proceeds step by step, method by method, from distances within our solar system to the largest cosmic distances of billions of light-years \citep{Grijs2011, Webb1999}. The fundamental concepts of this distance ladder are part of the curriculum of introductory astronomy courses \citep{CobleEtAl2013} and commonly considered basic astronomical knowledge -- essential for understanding how astronomers know what they know. 

\section{Standard candles}
Beyond the realm of nearby stars (with distances up to some tens of thousands of light-years), astronomical distance measurement relies on {\em standard candles}, as follows. Assume that a cosmic object emits the total energy $L$ per second. $L$ is called the object
s {\em luminosity}. Furthermore assume that the emission is isotropic: spread evenly over all directions, with no direction favoured or disfavoured.

Under such conditions, it is readily shown that the apparent brightness of the object, as seen by an observer at distance $d$, follows an inverse-square law. Let us consider the energy emission of the object as a serious of bursts, each of (very brief) duration $\Delta t$, seemlessly following each other. The energy $E=L\cdot\Delta t$ emitted in one such burst will be spread evenly on an expanding spherical shell. As it reaches the observer, it will have spread evenly over a spherical surface of radius $d$, that is, with area $4\pi d^2$. Hence, the fraction of energy $E_d$ that falls into a detector pointed directly at the cosmic object by the observer during the time $\Delta t$ will be proportional to the ratio of the detector's area $A$ and the spherical shell area, 
\begin{equation}
E_d = E \frac{A}{4\pi d^2} = L\cdot\Delta t\frac{A}{4\pi d^2}.
\end{equation}
Astronomers typically quote the energy flux $F$, that is, the energy detected per unit detector area per unit time. In our case, this means dividing the energy $E_d$ detected by the detection time and detector area, leading to the simple formula
\begin{equation}
F = \frac{L}{4\pi d^2}\label{InverseSquare}
\end{equation}
linking flux, luminosity and distance. Flux is, in principle, a measurable quantity.\footnote{In practice, the fact that each astronomical observation covers only a comparatively narrow range of the electromagnetic spectrum, and thus receives only a certain fraction of the total energy emitted by an object, introduces some complications, but while those make the analysis more involved, they do not change the fact that flux is something that can be measured by a distant observer.} A {\em standard candle} is defined as an object whose luminosity $L$ is known, or can be deduced from observable quantities. For such a standard candle, the measurable quantity $F$ and the known quantity $L$ can be used to solve equation (\ref{InverseSquare}) for the distance $d$: the apparent brightness (represented by the flux) and the known intrinsic brightness $L$ can be used to determine the object's distance.

Various standard candles are used in contemporary astronomy, using diverse physical laws or heuristic relationships to estimate luminosities. The first astronomical standard candles were the Cepheids:  variable stars whose period of variation (readily measured) is correlated with their luminosity.  First hints of this correlation were discovered by Henrietta Swan Leavitt in 1908, and a more precise formulation given by Leavitt in 1912. With the first calibration by Ejnar Hertzsprung in 1913, Cepheids became a distance indicator -- and, within the following years, they became instrumental in determining the first extragalactic distance scales, and in demonstrating that our galaxy is just one among many \citep{Fernie1969}.

Modern astronomy uses a variety of standard candles.  {\em Primary distance indicators} are those that can be calibrated within our home galaxy.\footnote{Alternatively, according to some authors: within our local group of galaxies, e.g. section 1.3.A in \citet{Weinberg2008}.} They include Cepheids, but also other types of variable stars with period-luminosity relations: RR Lyrae and Mira type variables. Other standard candles make use of luminosity-color diagrams (equivalently: luminosity-temperature diagrams) of star clusters: both for main sequence stars in the Hertzsprung-Russell diagram and for so-called red clump stars, this allows to deduce stellar luminosities within a fairly narrow range. Eclipsing binaries allow for another kind of estimate: a Doppler measurement of orbital velocities, combined with the light-curve data, leads to an estimate of the primary star's radius; this, when combined with the star's temperature (deduced from the spectrum) and the Stefan-Boltzmann law leads to an estimate of the star's luminosity\citep{Feast2005,Feast2013,Weinberg2008}.

{\em Secondary distance indicators} can be observed out to much greater distances, but need to be calibrated using distant galaxies for which primary distance indicators are known. A number of secondary standard candles use overall properties of galaxies. Examples are the Tully-Fisher relation, a heuristic correlation between a spiral galaxy's rotation speed (via measured via spectral line width) and stellar luminosity, the Faber-Jackson relation correlating elliptical galaxies' random stellar velocities (measured again via spectral line widths) with total luminosity, and the fundamental plane relation, a refinement of the Faber-Jackson relation that takes into account surface brightness as an additional observable parameter. The method of surface brightness fluctuations, on the other hand, makes use of the fact that a galaxy's stellar luminosity is made up of the light of myriads of individual stars; even at distances where individual stars can no longer be identified, this gives galaxies a slightly grainy appearance that varies with the distance between galaxy and observer \citep{Weinberg2008}.

The most famous secondary distance indicators are supernovae of Type Ia (SN Ia): thermonuclear explosions of white dwarf stars, for which there is a correlation between the time scale of the explosion (which can be read off of the supernova light curve) and its peak luminosity. Studies of the distance-redshift relation using these supernovae led to the discovery that the expansion of the universe is accelerating; a discovery that was honoured with the Nobel prize in physics in 2011 \citep{Schmidt2012}.

\section{Street lights as standard candles}
Typical classroom demonstrations of the inverse square law involve a photodetector, such as a small solar cell, placed at varying distances from a light source \citep{Stanger2008}. Here, we describe a different kind of hands-on activity, which treats street lights as standard candles. The activity uses a digital camera and freely available software. It is suitable as a high school class project as part of a physics course, or as an astronomy club activity.

The exercise works best along a long, straight road; for our worked-out example, we have chosen a straight road in Heidelberg, Germany that measures 660 m end to end. In the following, we will determine reference distances between the observer (represented in this case by an SLR camera) and the target objects (the street lights) using the distance measuring tool in Google Maps. This would be akin to testing a particular type of standard candle against another, well-established distance indicator.

\begin{figure}
\plotone{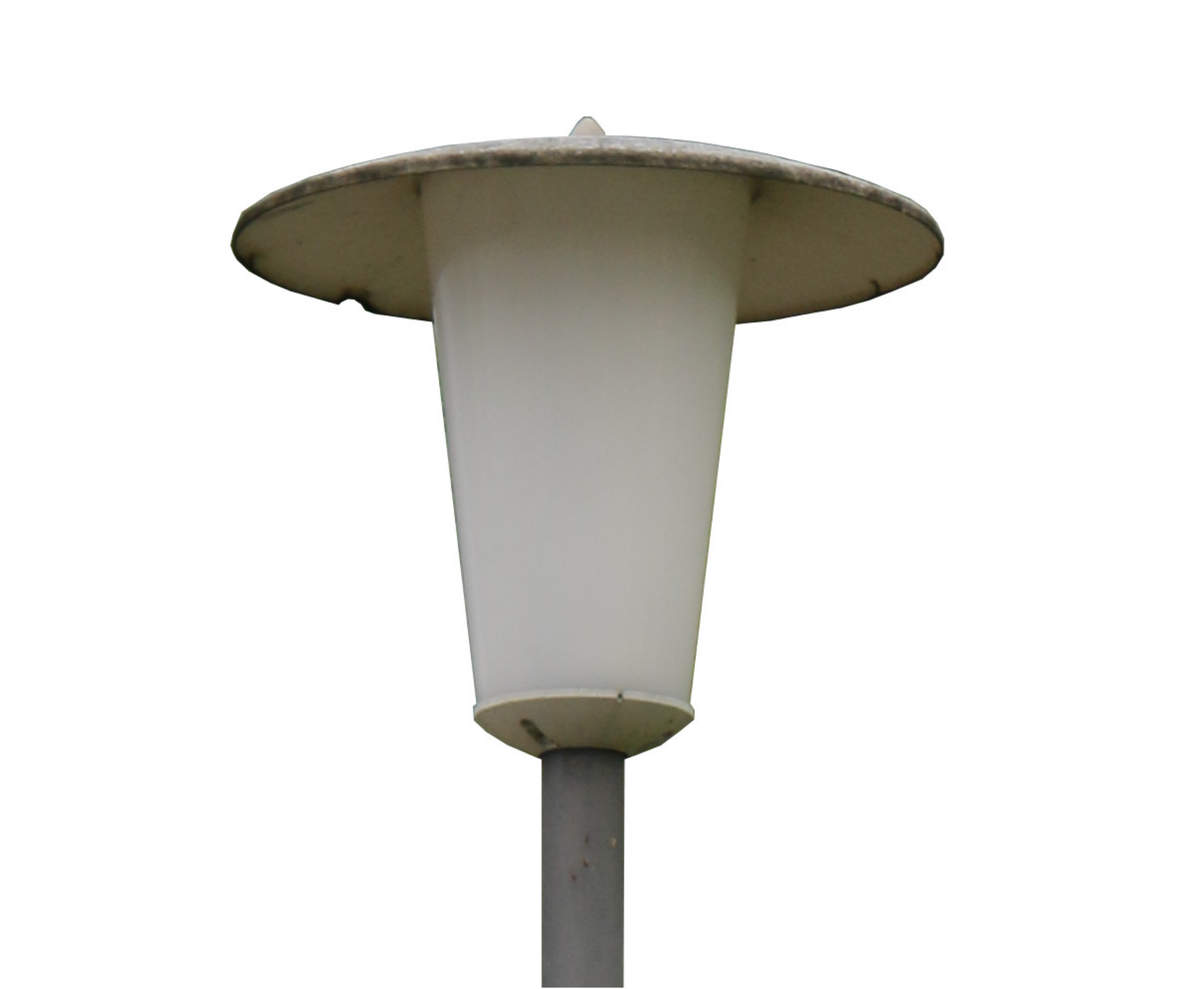}
\caption{Upper portion of a street light of the truncated-cone type used in the worked-out example \label{StreetLight}}
\end{figure}

Observations of street lights will usually be made from a lower height than that of the street light's lamp itself. For that reason, it is important that the street lights be of a type where the role of perspective is negligible. This rules out street lights whose light shines out downward from a horizontal surface; such a surface would present a larger section to an observer closer to and below the light than to a more distant observer, representing a systematic change in luminosity detrimental to standard candle measurements. Spherical street lights would be ideal; the street lights used here were of a truncated-cone type shown in fig. \ref{StreetLight}, which are fairly robust with respect to such changes of perspective. 

\begin{figure*}
\plotone{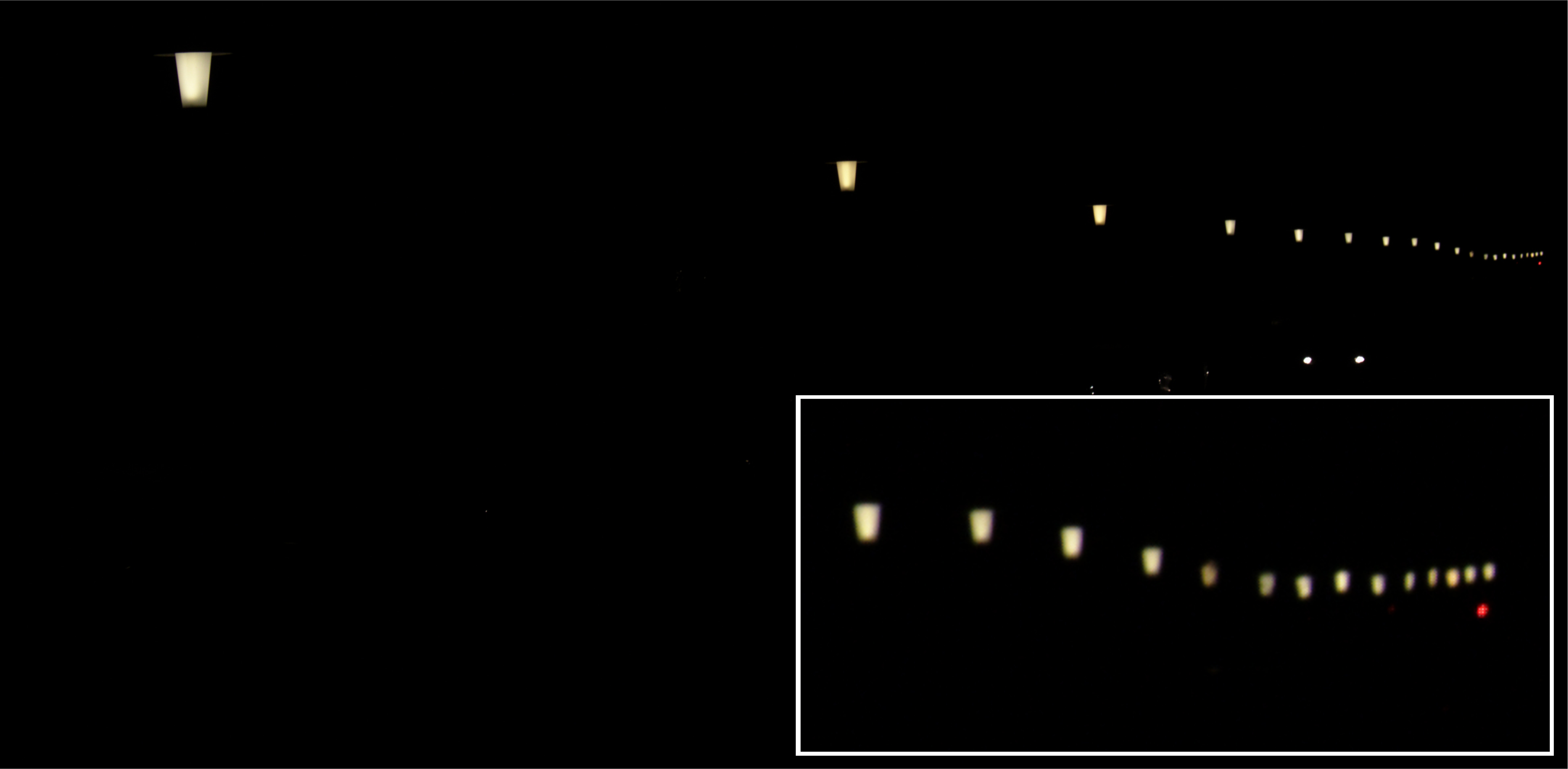}
\caption{Image of 20 street lights along a straight road at night. Inset bottom right: $\times 4$ zoom of the 14 most distant lights \label{PhotometricImage}}
\end{figure*}

Using a camera offset against the line of street lights at one end of the street, the line of lights can be captured in a single image, parts of which are shown in figure \ref{PhotometricImage}.

In order to read off the different street lamps' apparent brightness, pixel values within the image need to be linear -- if pixel A has received twice as much light as pixel B, pixel value A must be twice as large; pixel value must be proportional to the amount of light received by each pixel. For a DSLR camera, this is not a given. Images that are saved in JPG format, for instance, are processed in a non-linear fashion that is tailored to human expectations for images showing everyday situations. The images are meant to look natural, but they are not intended for quantitative measurements. Thus, the first challenge is to save the DSLR image in a way that allows for faithful measurements of brightness (in other words, for good photometry). 

The first step is to make the camera save the image in a raw format, which does not throw any of the brightness information away; the second is to transform the image into a format readable by software which can make simple brightness measurements: marking an image region and reading out the sum, or equivalently the mean, of all the pixel brightness values within that region. 

For astronomical photometry, there is a standard method called {\em aperture photometry}, which is of relevance for this situation. It amounts to summing up pixel brightness values within a region of area $A$ containing the object of interest, giving a total brightness $F_0$. This total brightness is due not only to light emitted by the object, but also to the background brightness of the sky contained within $A$; this background brightness is due in part to light scattered by the atmosphere, in part due to unresolved cosmic objects other than the object under scrutiny. The background contribution to $F_0$ can be estimated by measuring the average pixel value $p_{bg}$ within a surrounding annulus outside the area $A$. This makes $p_{bg}\cdot A$ a good estimate for the background brightness contribution to $F_0$, and
\begin{equation}
F_{obj} = F_0 - p_{bg}\cdot A
\label{AperturePhotometry}
\end{equation}
a good estimate for the brightness $F_{obj} $ due to the object of interest. Similarly, in the image showing the street lamps, it is necessary to estimate background brightness $p_{bg}$; we do so by measuring mean pixel brightness in a rectangular area directly above each lamp (hence shielded from the lamp's own light by the lamp shade). 

The resulting brightness values $F_{obj}$ using equation (\ref{AperturePhotometry}) are uncalibrated. As a simple way of calibration, we took the distance $d_{gm}(1)$ between the camera and the first street light, as measured using Google maps, as given; by the inverse square law (\ref{InverseSquare}), the distance of the $i$th street light should then be
\begin{equation}
d(i) = d_{gm}(1)\cdot\sqrt{\frac{F_{obj}(1)}{F_{obj}(i)}}.
\label{StreetLightSC1}
\end{equation}

\begin{figure}
\plotone{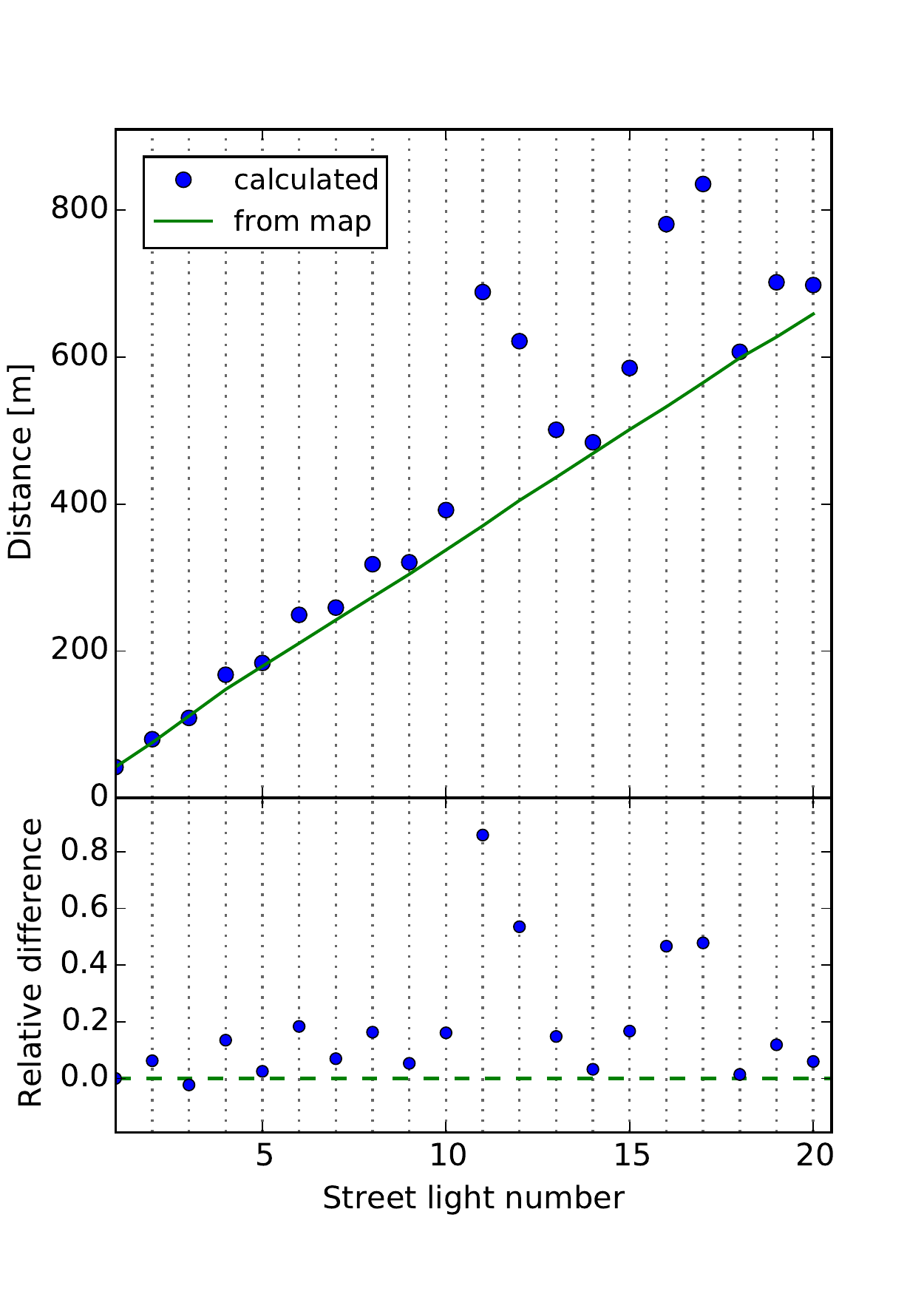}
\caption{Top: Distances computed with a straightforward application of the standard candle ($L=const.$) inverse square law (blue dots), compared with Google Maps distances (green line). Bottom: Relative deviations between calculated distances and map distances.  }
\label{FirstDistancePlot}
\end{figure}
The upper plot in figure \ref{FirstDistancePlot} shows the distances inferred by treating the street lights as standard candles, using equation (\ref{StreetLightSC1}). Both these distances $d(i)$ and the Google map distances $d_{gm}(i)$ are plotted against street light number. The lower plot shows the relative difference $[d(i)-d_{gm}(i)]/d_{gm}(i)$ between the two. Ideally, the two distance values should be the same for each street light. Yet the plot shows marked discrepancies for a few of the lights (in particular numbers 11, 12, 16 and 17). What has gone wrong?

\section{Standard candle problems}

Using street lights as standard candles leads to problems similar to those involving astronomical standard candles in general; in other words: from analysing problems with their street light measurements, pupils can learn about confounding factors that occur in analyses of astronomical standard candles, as well.

Eq. (\ref{StreetLightSC1}) assumes that all street lights have the same luminosity -- but what if there are slight (or even not so slight) variations? In those cases, distance values calculated via that equation will be skewed systematically -- street lights that are intrinsically fainter will appear as further away than they really are; abnormally bright street lights will appear as significantly closer.

This is a key problem for astronomical standard candles: How homogeneous is a population of objects? Do the objects under consideration indeed all have the same luminosities, or are there variations -- and if there are, are there measurable properties correlated with these variations, which can be used to predict the luminosity values for each of the objects? The most famous example of a standard candle that underwent this kind of correction are, once more Supernovae of Type Ia: Initially, they were thought to be standard candles with approximately constant luminosity. But work in the early 1990s showed that there is a correlation between the time scale of a supernova's light curve and its peak luminosity: light curves that decline fastest are characteristic of the dimmest supernova explosions \citep{Phillips1993}.

\begin{figure}
\plotone{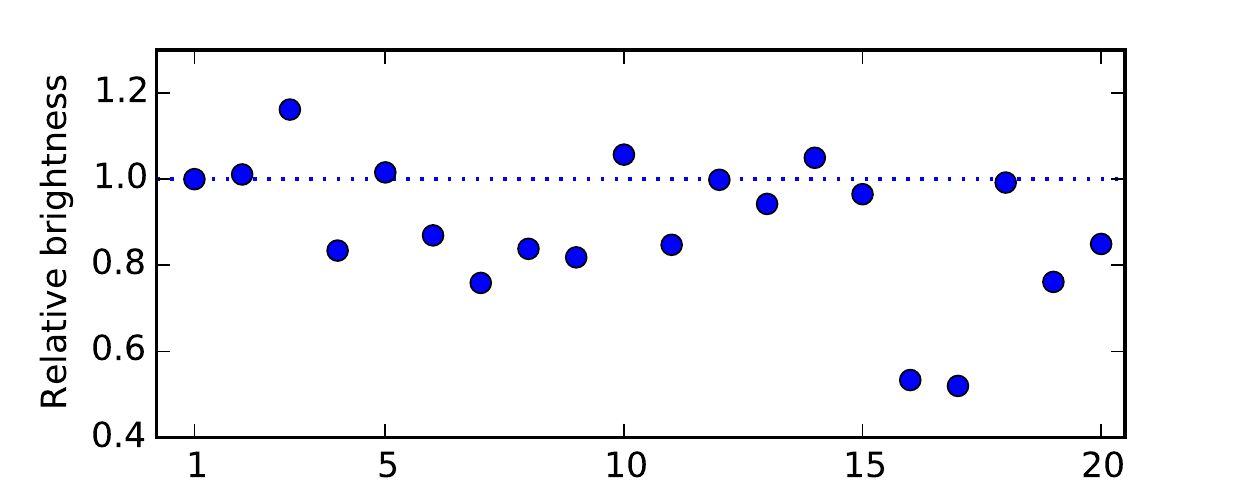}
\caption{Luminosities of the various streetlights relative to the first streetlight, as determine by observing each from a standard distance.}
\label{relativeLuminosities}
\end{figure}

In the case of street lights, it will hardly be worth our while to formulate intricate physical models, but instead we have the advantage of being able to document brightness variations directly. The proper way of doing this would be to take an image of each street light from a standard distance and, to account for emission anisotropies, from the same direction as for the image used to measure relative street light brightnesses. Instead, we have followed a simplified procedure, taking an image of each street light from an approximately standardised distance (not measured precisely, but counted off in steps) in the direction along the street (which deviates from our viewing direction for the brightness measurements by $14^{\circ}$ for the first light, rapidly decreasing to less than $4^{\circ}$ by the time we have reached the 4th light). Luminosities relative to the first light,
\begin{equation}
r(i) \equiv \frac{L(i)}{L(1)}
\end{equation}
for $L(i)$ the luminosity of the $i$th street light,  as estimated by that method, are shown in fig. \ref{relativeLuminosities}. 

\begin{figure}
\plotone{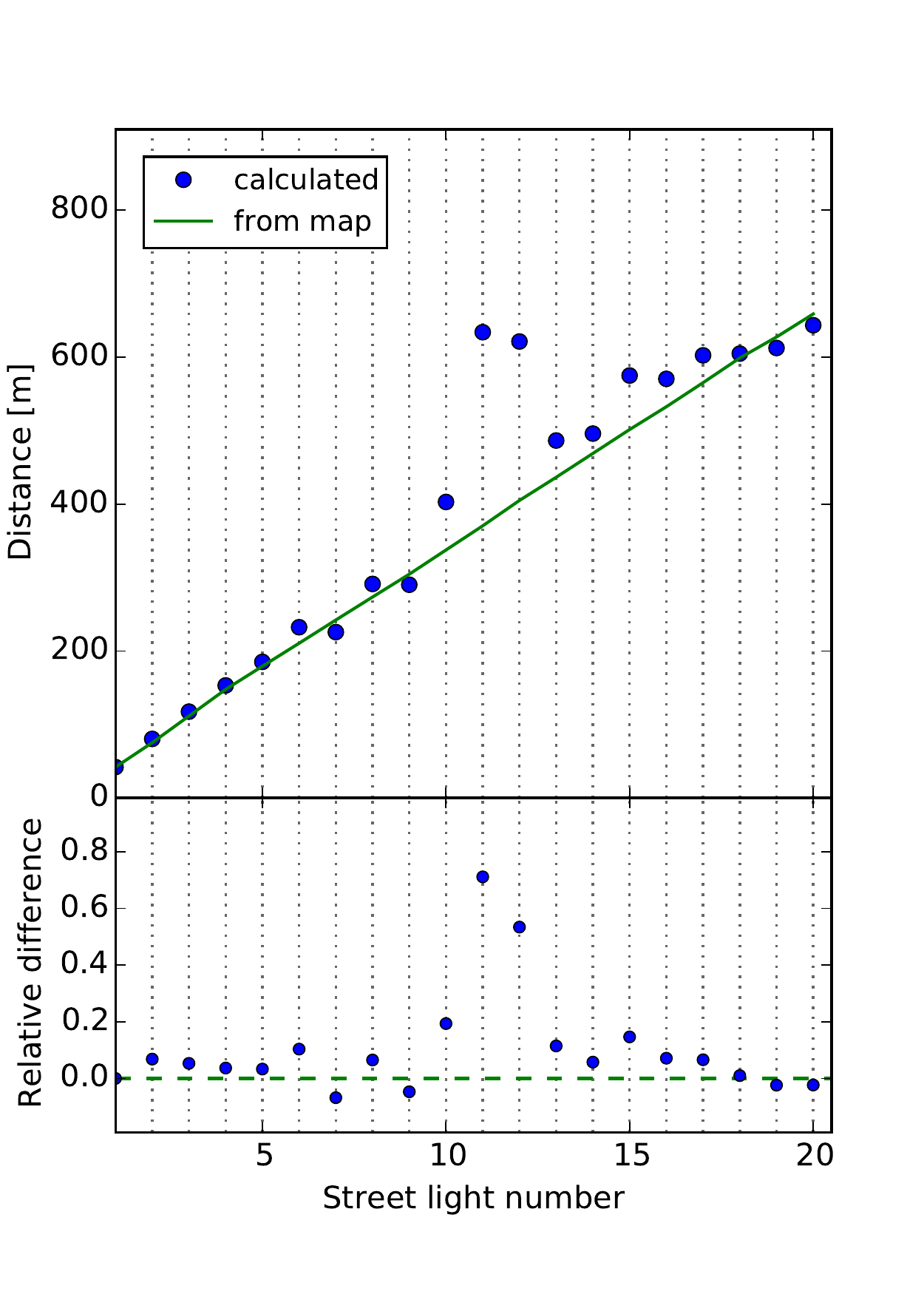}
\caption{Top: Distances computed via the inverse square law taking into account individual luminosity variations (blue dots), compared with Google Maps distances (green line). Bottom: Relative deviations between calculated distances and map distances.}
\label{LuminosityCorrectedDistancePlot}
\end{figure}

Taking into account such intrinsic variations of luminosity, the new formula linking distance and apparent brightness is
\begin{equation}
d(i) = d_{gm}(1)\cdot\sqrt{r(i)\cdot \frac{F_{obj}(1)}{F_{obj}(i)}}.
\label{StreetLightSC2}
\end{equation}
The upper plot in fig. \ref{LuminosityCorrectedDistancePlot} shows, for each street light, the standard candle distance calculated in this way as well as the Google map distance; the relative differences of the two are once more shown in the lower plot.

This has taken care of the outliers 16 and 17; direct inspection of those two street lights shows that they have been shielded so as not to disturb inhabitants of the nearby houses which, at that point of the street, have windows rather close to those lights. What, then, is the remaining problem with lanterns 11 and 12?

An image taken at greater magnification reveals another problem that has an analogue for astronomical standard candles: {\em extinction}, that is, the absorption/scattering of light as it passes through intervening matter. 

In astronomy, extinction is typically due to gas and dust that are located between the observer and the target object. Astronomical extinction can be estimated making use of the fact that different wavelengths of light are affected by extinction in different ways.

\begin{figure}
\plotone{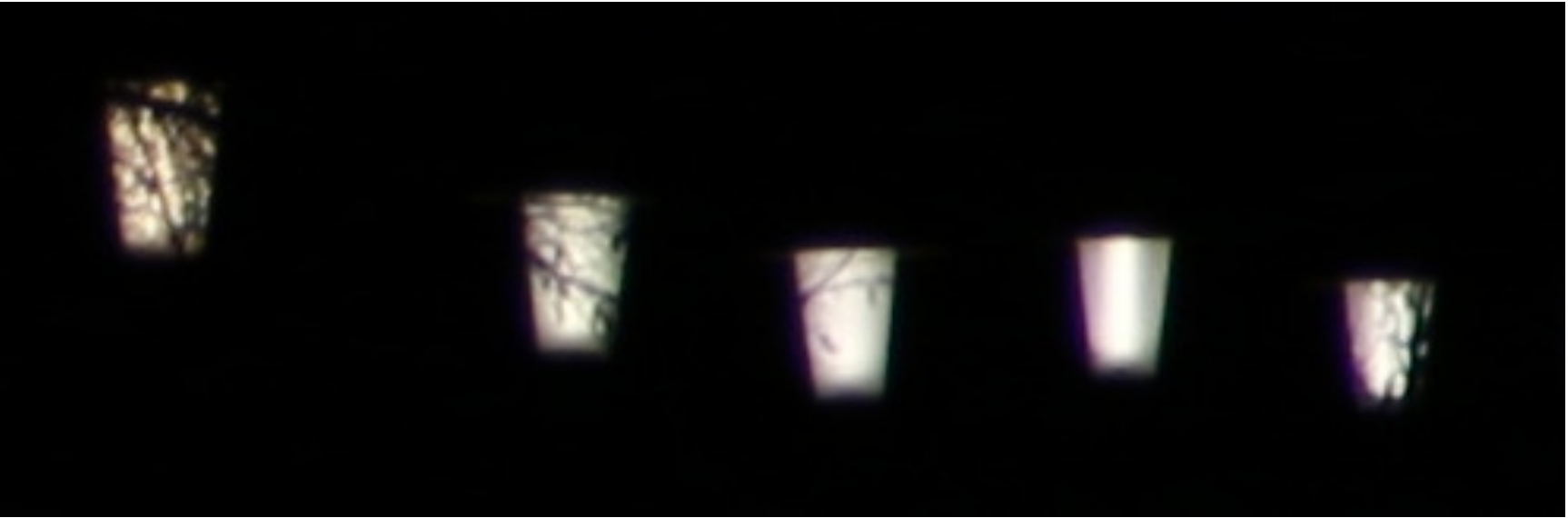}
\caption{Image showing, left to right, street lights 11 to 15. Apart from street light 14, apparent brightness is visibly affected by tree branches between light and observer
\label{TreeShading}}
\end{figure}

In our case, taking an image of the more distant street lights directly shows extinction due to shading by branches from overhanging trees, as can be seen in fig. \ref{TreeShading} for the street lights 11, 12, 13 and 15. Going by direct visual inspection of the image, those four lights are the only ones affected.

The fraction of the total area shaded by the branches can be estimated in the following simple way: Measure the mean pixel brightness $p_{lamp}$ for the whole street lamp, including the obscured parts. Measure the mean pixel brightness for selected parts of the lamp that are not obscured, and use their average (weighted with area) to estimate the lamp's unobscured mean brightness $p_{unobs}$. Then
\begin{equation}
f = 1-\frac{p_{lamp}}{p_{unobs}}
\end{equation}
yields a good estimate for the fraction of lamp light lost due to obscuration. Values for the four affected lamps are given in table \ref{ObscurationTable}.

\begin{deluxetable}{cc}
\tablecaption{Obscuration coefficients (percentage of light held back) for four affected street lamps\label{ObscurationTable}}
\tablehead{\colhead{Streetlight no. $i$} & \colhead{f(i)} }
\startdata
11 & 57.4\%\\[0.2em]
12 & 48.9\%\\[0.2em]
13 & 21.6\%\\[0.2em]
15 & 42.6\%
\enddata
\end{deluxetable}

\begin{figure}
\plotone{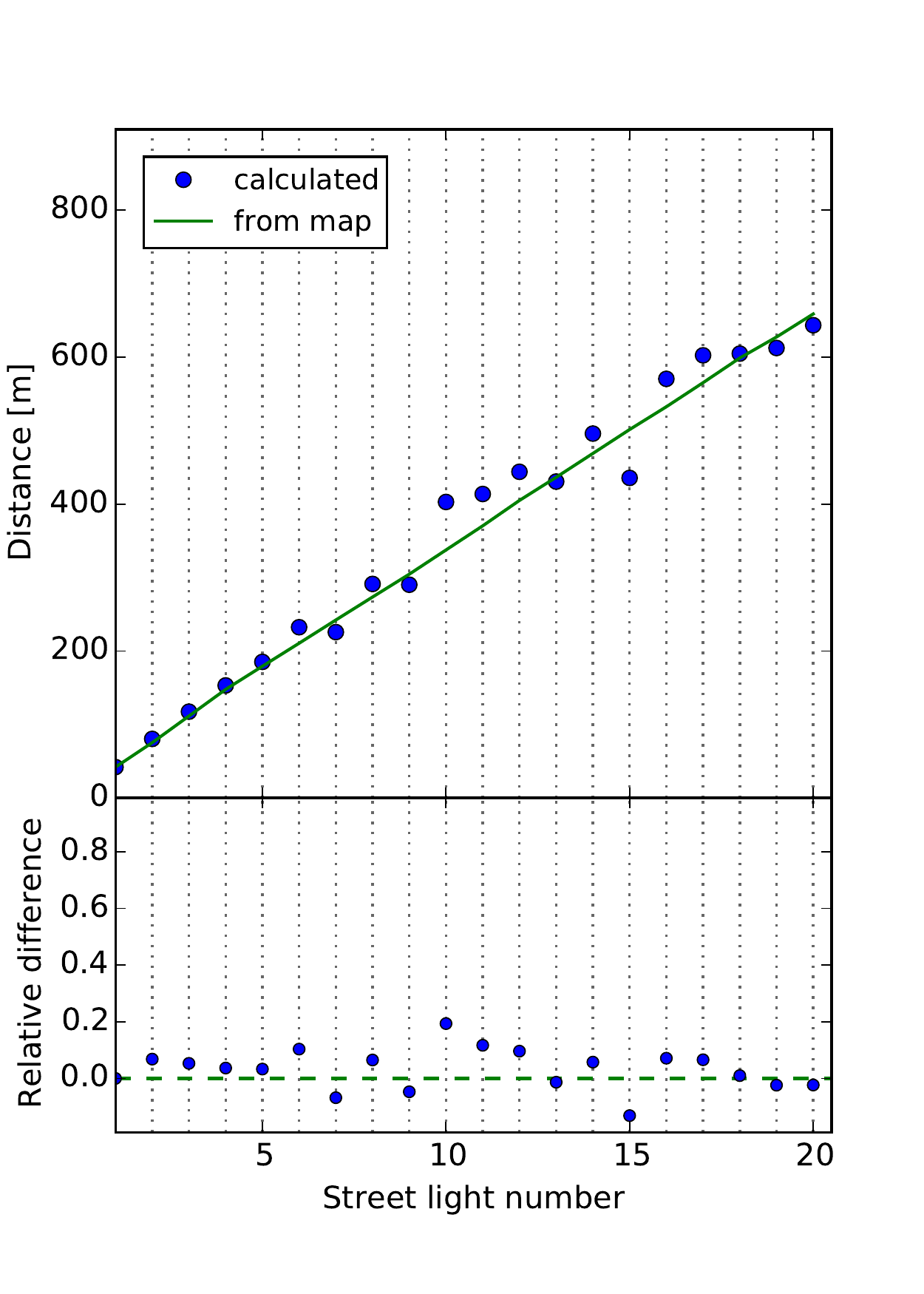}
\caption{Top: Distances computed via the inverse square law taking into account individual luminosity variations and extinction (blue dots), compared with Google Maps distances (green line). Bottom: Relative deviations between calculated distances and map distsances.\label{FinalCorrectedDistances}}
\end{figure}

In our case, extinction means that we need to multiply the luminosity $L(i)$ of the $i$th street light by $1-f(i)$ in order to calculate the expected amount of light received. The modified standard candle equation reads
\begin{equation}
d(i) = d_{gm}(1)\cdot\sqrt{r(i)\cdot\frac{1-f(i)}{1-f(1)}\cdot \frac{F_{obj}(1)}{F_{obj}(i)}}.
\label{StreetLightSC3}
\end{equation}
The result, once more in comparison with the Google Map distances, is shown in fig. \ref{FinalCorrectedDistances}. 
For more than two thirds of the street lights, the agreement is now better than 10\%; in astronomy, agreement around the 10\% mark is commonly considered a good (or even great) result. 

There is still a tendency of calculated inverse-square law distances to lie above the green line, hinting at systematic errors. The simplest such error would be a problem with the calibration which, after all, relies on a single Google Maps measurement (namely that of the distance between the camera location and the first street light) and on estimates for the street lanterns' relative luminosities.

The set-up described here is suitable for a basic in-the-field demonstration of inverse square law distance determinations. Where higher precision is desired, a number of opportunities for improvements suggest themselves: Instead of comparing the distances between street lights with an online map, we could directly determine those distances using tape measures. For the luminosities, we could ensure a more strictly standardised measurement by taking into account the exact perspective of the camera to suppress effects of direction-dependent lighting. For the extinction estimate, we could use images taken at an even greater focus length. 

\section{Conclusions}

Street lights as standard candles provide a useful hands-on activity that demonstrates to the learners the basic principles of distance measurement via standard candles. In contrast with laboratory measurements where one varies the distance between one light source and the observer/detector, the activity can also serve to illustrate some of the main potential problems with standard candle observations: unforeseen variations of standard candle luminosity and extinction effects. 

The tools used, namely a DSLR and simple image processing software, are similar to their professional astronomical counterparts as to make the activity a suitable first introduction to astronomical photometry, and they have a good chance of being available in a high school setting. Time spent getting to know these tools is well invested, as they can be used for numerous types of direct astronomical observation, as well.
Data analysis is simple, using variations of the inverse square law, modified by ratios of physical quantities. The calculations themselves are sufficiently simple as to be suitable for a Microsoft Excel or Libreoffice Calc spreadsheet.

\begin{appendix}
\section{Supplementary information for the worked-out example}

The street chosen was {\em Albert-Fritz-Stra\ss e} in Heidelberg, Germany; the beginning of the street can be found on Google Maps via the link [\href{http://www.google.com/maps/@49.382752,8.663361,19z}{http://www.google.com/maps/@49.382752,8.663361,19z}]. In this view, the street lights are on the upper pavement; numbers in the text have been assigned from left to right, starting with the street's second light. Measurements can be made using Google's measuring tool in the classic map mode and may require the user to be logged in. Remember to uncheck the ``45$^{\circ}$'' option for a straight-down view. For the image, the camera was positioned right below the top left corner of the roof of first house in the street, on the lower side in the standard Google view. The lanterns are visible as white circles on the upper pavement (and cast shadows toward the top right). 

Images and supplementary material (including an Excel worksheet and ImageJ macros documenting the photometric measurements) can be found on [\href{http://www.haus-der-astronomie.de/materials/distances/street-lights}{http://www.haus-der-astronomie.de/materials/distances/street-lights}].

The photometric image for this example was taken with a Canon EOS 70D with 18--55 mm Canon kit lens at 55 mm, exposure 1/160 s at f/8.0 and ISO 200. Lower ISO values make for lower image noise; the f stop should be chosen in a way to allow for the street lights at their different distances all to be in focus, and exposure should be such that no pixel is saturated. (A quick way of checking is to plot a profile of pixels across the brightest lamp in the image; if that profile has a flat top region, there is likely to be saturation.)

The raw (.CR2) files from the camera were transformed into FITS files using Fitswork [\href{http://www.fitswork.de}{http://www.fitswork.de}] (which in turn uses dcraw [\href{http://www.cybercom.net/~dcoffin/dcraw/}{http://www.cybercom.net/\~{}dcoffin/dcraw/}] for the conversion), and brightness measurements were made with the polygon tool in ImageJ [\href{http://imagej.nih.gov/ij/}{http://imagej.nih.gov/ij/}]. All brightness measurements made in ImageJ were recorded as macros, for reference; the macro files are available as part of the supplementary material available for download.

ImageJ measurements can be saved as tab-separated files (which, confusingly, will get a .xls extension; when opened with Microsoft Excel or LibreOffice Calc, these files will need to be imported; if you are in a country that uses decimal commas instead of decimal points, make sure that decimal places get imported correctly).

The image showing extinction by tree branches was taken with the same Canon EOS 70D, using a Tamron 18-200 mm zoom lens at 200 mm.

\section{Curriculum connections}

Standard candles and the inverse-square law are key concepts in astronomy. Both concepts are necessary to understand modern determination of distances beyond our Solar System; such distance measurements, in turn, are a necessary prerequisite to deducing certain physical properties - most importantly intrinsic luminosity - from observations of astronomical objects. 

In the UK, the astronomy GCSE qualifications prepared by EdExcel \citep{Gregory2011}, which complements the Key Stage 4 (KS4) science curriculum (GSCE Science and GSCE Additional Science) include the understanding of apparent and absolute magnitude, their relation to distance, the inverse-square law and its use with Cepheid standard candles in Topic 3.3. ``Physical Properties of Stars'' of Unit 1 ``Understanding the Universe'' (5AS01). Skills obtained in the activity described above, namely photometric measurements using a digital camera and image-processing software for analysis, can be re-used for some of the observation activities in Unit 2 ``Exploring the Universe''  (e.g. observation tasks B7, B9).

The Scottish ``Curriculum for Excellence'' lists standard candles as part of their higher physics curriculum area ``Our Dynamic Universe'': Under the heading ``The expanding Universe'', the Course Unit Support Notes explicitly list ``Consideration of parallax measurement and the data analysis of apparent brightness of standard candles in measuring distances to distant objects'' \citep{SQA2014}.

In Germany, astronomy is a mandatory part of the high school curriculum in some of the country's federal states. In \cite{MV2004}, the basic understanding of astronomy (section 7.1) requires understanding of light as a carrier of information, and of receivers such as photographic cameras or CCDs; mandatory content for understanding stars and stellar systems (section 7.6) includes brightness and distances of the stars, methods of distance determination, and the concept of absolute brightness. Astrophotography is a suggested project activity (appendix 3).

In Sachsen-Anhalt (Saxony-Anhalt), the obligatory 9th grade curriculum for high schools (Gymnasium) includes a segment ``Stars and star system'' (Thema 5), part of which is an experimental demonstration of the relationship between distance and apparent brightness \citep{SA2003}. The goals of the general 10th grade astronomy curriculum include an understanding of the relationship between luminosity, apparent brightness and distance  \citep{SA2012}. For grades 11/12, the curriculum for astronomy courses includes (Kurs 2) the relationship between apparent magnitude, absolute magnitude, and brightness \citep{SA2003}.

In Th\"uringen (Thuringia), 10th grade astronomy includes an understanding of the difference between apparent and true brightness as well as of at least one method of determining astronomical distances. Methods for determining astronomical distances, including photometric distances and different types of standard candle, are part of the 12th grade curriculum (Themenbereich 3.2). More generally (Themenbereich 3.1), students are expected to be competent in making their own astronomical observations, including the use of digital photography and software for image processing and analysis \citep{Thu2012}.

In Bavaria, astrophysics is one of two possible themes for teaching physics in grade 12, the other being quantum physics \citep{ISB2004}. In the section on stars (Ph$_{\mathrm Ast}$ 12.4), students are required to learn about apparent and absolute magnitudes as well as the distance modulus; in the section on large-scale cosmic structure (Ph$_{\mathrm Ast}$ 12.5), about standard candles, notably Cepheids and supernovae. 
\end{appendix}

\end{document}